\documentclass[showpacs,amssymb,preprint,preprintnumbers,nofootinbib,superscriptaddress]{revtex4}

\usepackage{amsmath}
\usepackage{graphicx}
\graphicspath{{Figures/}}
\usepackage{latexsym}
\usepackage{amsfonts}
\usepackage{url,hyperref}
\usepackage{bm}
\usepackage{textcomp}
\usepackage{color}
\usepackage{bbm}
\usepackage{slashed}
\usepackage{caption}
\usepackage{epstopdf}
\usepackage{subcaption}
\usepackage{placeins}
\usepackage{color}

\newcommand{\beq}{\begin{equation}}
\newcommand{\eeq}{\end{equation}}
\def\bea{\begin{eqnarray}}
\def\eea{\end{eqnarray}}

\begin{document}

\title{A new (original) set of Quasi-normal modes in spherically symmetric AdS black hole spacetimes}

\author{Chun-Hung~Chen}
\email[Email: ]{chun-hungc@nu.ac.th}
\affiliation{The Institute for Fundamental Study, Naresuan University, Phitsanulok 65000, Thailand.}

\author{Hing-Tong~Cho}
\email[Email: ]{htcho@mail.tku.edu.tw}
\affiliation{Department of Physics, Tamkang University, Tamsui District, New Taipei City, Taiwan 25137.}

\author{Alan~S.~Cornell}
\email[Email: ]{acornell@uj.ac.za}
\affiliation{Department of Physics, University of Johannesburg, PO Box 524, Auckland Park 2006, South Africa.}


\begin{abstract}
From black hole perturbation theory, quasi-normal modes (QNMs) in spherically symmetric AdS black hole spacetimes are usually studied with the Horowitz and Hubeny methods \cite{hh2000} by imposing the Dirichlet or vanishing energy flux boundary conditions. This method was constructed using the scalar perturbation case and {\it box-like} effective potentials, where the radial equation tends to go to infinity when the radial coordinate approaches infinity. These QNMs can be realized as a different set of solutions from those obtained by the {\it barrier-like} effective potentials. However, in some cases the existence of {\it barrier-like} effective potentials in AdS black hole spacetimes can be found. In these cases this means that we would obtain a new (original) set of QNMs by the purely ingoing and purely outgoing boundary conditions when the radial coordinate goes to the event horizon and infinity, respectively. Obtaining this set of QNMs in AdS black hole cases is the main focus of this paper.
\end{abstract}


\date{\today}

\maketitle

%
%
\section{Introduction}

\par With the recent progress of black hole astronomy, including the detection of gravitational waves, the multi-messenger of binary neutron star mergers, and the first astronomical image of a supermassive black hole, the phenomena around black holes are expected to be further tested in the coming future. Quasi-normal modes (QNMs) corresponding to various spin fields which perturb black hole spacetimes are in the catalogue of processes which dominate the ring-down phases of the wave propagation from such observed objects. The four dimensional and asymptotically de-Sitter (dS) spacetime cases may be the main focus of recent experimental results, however, in string theory and within the AdS/CFT correspondence, black hole perturbation theory in higher dimensional asymptotically Anti-de Sitter (AdS) spacetimes still play an important role. The most explicit result being the QNMs of asymptotically AdS spacetimes, which correspond to the poles of the Green's functions in the corresponding conformal field theory (CFT) \cite{hh2000, cl2001, ckl2003, wlm2004, gj2005}. In spherically symmetric black hole spacetimes the QNMs are obtained by the same radial equations as the cases representing the ring-down behavior, and also the asymptotic AdS ones, which shall be of the Schr\"odinger-like form:
\begin{equation}\label{RD}
\frac{d^{2}}{dr_{\ast}^{2}}\Psi_{s}+\left(\omega^{2}-V_{s}\right)\Psi_{s}=0 \; ,
\end{equation}
where $r_{\ast}$ represents the tortoise coordinate defined as $dr_{\ast}=f^{-1}\left(r\right)dr$ in most  massless perturbation cases, except the spin-3/2 one which will be discussed later. The $V_{s}$ represents the effective potential, dominated by the behavior of the propagation of waves, where $s$ represents the spin of the perturbating fields. The master equations in this form allow us to study the characteristics of the propagating waves by analogies to well studied quantum mechanics methods.

\par Our focus in this paper shall be the asymptotically AdS cases, where the first study of QNMs in general dimensional spherically symmetric AdS black hole spacetimes was done by Horowitz and Hubeny for scalar perturbations which satisfied the Klein Gordon equation in curved spacetimes \cite{hh2000}. The study started with the well known Regge-Wheeler equation, then by performing a systematic transformation to the radial equation, and the imposition of the Dirichlet boundary condition, the QNMs were obtained. Several independent groups have generalized this method to the Dirac, electromagnetic and gravitational perturbations \cite{cl2001, ckl2003, wlm2004, gj2005}, as well as considering a more generic boundary condition with a vanishing energy flux \cite{whs2015, whj2017}.
The papers in Refs. \cite{hh2000, cl2001, ckl2003, wlm2004, gj2005} have further indicated a relationship, under the AdS/CFT correspondence, between the QNMs and the poles of the Green's functions in the CFT. As such, the general expectation of the asymptotically AdS spacetimes is that the effective potential, $V_{s}$, tends to go to infinity when the spatial coordinate $r\rightarrow\infty$, which is realized with {\it box-like} effective potentials. The {\it box-like} behavior naturally fits under the AdS/CFT correspondence for the Dirichlet boundary condition.

\par However, in our recent work on spin-3/2 perturbations in general dimensional Schwarzschild AdS ($SAdS_{D}$) spacetimes \cite{ccch2019}, in the four dimensional case it was possible to have {\it barrier-like} effective potentials in the cases of small black holes. This is unexpected of the general behavior of asymptotically AdS black hole spacetimes. With further comparison to other cases, we found that for several fermionic and bosonic perturbations, in four and higher dimensions, there were {\it barrier-like} effective potentials. The phenomena of {\it barrier-like} effective potentials suggested to us that mixed boundary conditions should also be considered when the wave propagates to the spatial infinity.

\par The AdS spacetimes fail to be globally hyperbolic as the propagation of a classical wave with a well-defined dynamic is possible if and only if suitable boundary conditions are imposed. A study of the feasibility of various kinds of boundary conditions for bosonic perturbations in general dimensional  AdS spacetimes was done by Ishibashi and Wald \cite{IW2004}. They classified their argument into four typical cases, where some of these cases were allowed to impose the general Robin boundary conditions (that is, beyond the Dirichlet and the vanishing energy flux boundary conditions). We found that the asymptotic behavior for our cases, for bosonic perturbations or one step further for fermionic perturbations, fit their classifications, and the presence of the black hole implies a barrier-like effective potential which is similar to those in the usual black hole perturbation studies. As such, the imposing of boundary conditions with purely ingoing and purely outgoing waves (when the radial coordinate goes to the event horizon and infinity, respectively), shall be naturally reasonable for these {\it barrier-like} effective potentials, as this boundary condition is equivalent to a specific choice of the general Robin boundary conditions.

\par We certainly agree that the Dirichlet boundary conditions have many explicit and well-studied physical interpretations under the AdS/CFT correspondence. However, this should not restrict us from considering other sets of boundary conditions in asymptotic AdS spacetimes because the Dirichlet boundary conditions are actually not appropriate to describe physical phenomena in some specific models. An explicit example, as presented in Refs. \cite{CM2008, H2009, AST2009}, shows that in order to have dynamical gravitational degrees of freedom on the AdS boundary, the Dirichlet boundary conditions should not be adopted. More specifically, when we consider a cosmological model with the physical brane imbedded in a $SAdS_{D}$ bulk spacetime, the mixed boundary conditions are necessarily imposed to have cosmological evolutions on the brane as it approaches the AdS boundary. An example of this was done in a holographic study for a four-dimensional Robertson-Walker brane imbedded in $SAdS_{5}$ in Ref. \cite{AST2009}. These works motivate us that the studies of our current set of QNMs, which are the formal solutions for the linearized radial equations in the $SAdS_{D}$ spacetimes but without imposing the {\it box-like} boundary conditions, should possess physical interpretations in such research directions and therefore warrant further study.

\par In the next section, we follow the well-studied radial equations corresponding to various bosonic and fermionic fields, studying the asymptotic behavior of the effective potentials, collecting the {\it non-box-like} and {\it barrier-like} behaviors and then comparing to the analogous Ishibashi and Wald classifications \cite{IW2004}. In section 3, we evaluate the QNMs with the outgoing wave boundary condition at spatial infinity, by using the revised WKB methods with Pad\`e approximants provided in Ref. \cite{KZZ2019}. We compare our modes with some reference modes obtained from the {\it box-like} boundary conditions, and show that our QNMs are a new set of solutions in Schwarzschild AdS black hole spacetimes. Lastly, we will present our discussions and some possible future research directions for this new set of QNMs.

%
%
\section{The {\it non-box-like} effective potentials for perturbations in $SAdS_{D}$ spacetimes}

\par The metric of $D=\left(n+2\right)$-dimensional $SAdS_{D}$ spacetimes is given by
\begin{equation}
ds^{2}=-f\left(r\right)dt^{2}+f^{-1}\left(r\right)+r^{2}d\Omega_{n}^{2} \; ,
\end{equation}
where
\begin{equation}\label{f}
f\left(r\right)=1-\frac{2M}{r^{n-1}}+\frac{2\Lambda r^{2}}{n\left(n+1\right)} \; ,
\end{equation}
$d\Omega_{n}^{2}$ is the metric on $S^{n}$, and the cosmological constant $\Lambda$ is always positive for the asymptotically AdS cases in this notation. By considering various types of perturbations, the necessary condition for {\it non-box-like} effective potentials is to be asymptotically non-infinite. Next, setting the parameters of the black hole mass $M$, and the cosmological constant $\Lambda$, one may obtained the {\it barrier-like} potential. We will present this characteristic for various effective potentials case by case in the following subsections. Note that in this paper we shall focus only on the massless perturbation cases.

%
\subsection{The scalar perturbation}

\par The effective potential for scalar perturbations in Eq.~(\ref{RD}) is given by \cite{BCO2009}
\begin{eqnarray}
V_{s=0}&=&f\left(r\right)\left[\frac{l\left(l+D-3\right)}{r^{2}}+\frac{\left(D-2\right)\left(D-4\right)}{4r^{2}}f\left(r\right)-\frac{\left(D-2\right)}{2r}\frac{df\left(r\right)}{dr}\right]\nonumber\\
&=&f\left(r\right)\left[\frac{l\left(l+D-3\right)}{r^{2}}+\frac{\left(D-2\right)\left(D-4\right)}{4r^{2}}+\frac{\left(D-2\right)^{2}}{4}\frac{2M}{r^{D-3}}+ \frac{D \Lambda}{2\left(D-1\right)}\right]\; ,
\end{eqnarray}
where $l=0,\ 1,\ 2,\ ...$ corresponds to the the scalar spherical harmonics. The asymptotic behavior when $r\rightarrow\infty$ is
\begin{equation}
V_{s=0}\Big|_{r\rightarrow\infty}\sim f\left(r\right)\frac{D \Lambda}{2\left(D-1\right)} \sim \frac{D \Lambda^{2} r^{2}}{\left(D-2\right)\left(D-1\right)^{2}} \; .
\end{equation}
It is obvious that the effective potential tends to go to infinity in this limit, and all the modes with different $l$ satisfy the general expectation of the asymptotically AdS cases, and the effective potential is always {\it box-like}.

%
\subsection{The Dirac perturbation}

\par The effective potential for Dirac perturbation in Eq.~({\ref{RD}}) is given by \cite{ccdn2007}
\begin{eqnarray}
V_{s=1/2}= \pm f\left(r\right)\frac{d W}{dr}+W^{2}\; ; \; \; W=\frac{\sqrt{f}}{r}\left(l+\frac{D-2}{2}\right)\; ,
\end{eqnarray}
where the sign $\pm$ represents a pair of supersymmetric partner potentials, and $l=0,\ 1,\ 2,\ ....$ corresponds to the eigenspinor on the sphere. Taking the positive sign, one can further simplify $V_{s=1/2}$ as
\begin{eqnarray}
V_{s=1/2}=W\left(\frac{1}{2}\frac{df\left(r\right)}{dr}-\frac{f\left(r\right)}{r}+W\right) \; .
\end{eqnarray}
One can check that the highest order of the radial dependance, $r$, for the super potential $W\sim r^{0}$, and the $\Lambda$ dependance term, is equivalent to the ${\cal O} (r^{1})$ term, for the first two terms inside the parentheses coincidentally cancel out. We can summarize the asymptotic behavior when $r\rightarrow\infty$ for the Dirac perturbation as
\begin{equation}
V_{s=1/2}\Big|_{r\rightarrow\infty}\sim\frac{2\Lambda}{\left(D-1\right)\left(D-2\right)}\left(l+\frac{D-2}{2}\right)^{2} \; ,
\end{equation}
which is a finite non-zero constant. This indicates that the effective potentials for Dirac perturbations in $SAdS_{D}$ have {\it non-box-like} behavior in the general dimensional cases.

%
\subsection{The electromagnetic perturbation}

\par For the electromagnetic perturbation we follow the work done by Crispino, Higuchi and Matas \cite{chm2001}, and take the {\it physical modes} into consideration. The physical modes generated by the scalar spherical harmonic and vector spherical harmonic arise when separating the angular part in each case. This is why the radial equations were known as EM-scalar perturbations and EM-vector perturbations.

\subsubsection{The EM-scalar perturbation}

\par The effective potential in this case is given by
\begin{eqnarray}\label{vems}
V_{s=1,S}&=&f\left(r\right)\left[\frac{l\left(l+D-3\right)}{r^{2}}+\frac{\left(D-2\right)\left(D-4\right)}{4r^{2}}f\left(r\right)-\frac{\left(D-4\right)}{2r}\frac{df\left(r\right)}{dr}\right],\nonumber\\
&=&f\left(r\right)\left[\frac{l\left(l+D-3\right)}{r^{2}}+\frac{\left(D-2\right)\left(D-4\right)}{4r^{2}} \right. \nonumber \\
&& \hspace{3cm} \left. -\frac{\left(3D-8\right)\left(D-4\right)M}{2r^{D-1}}+\frac{\left(D-4\right)\left(D-6\right)\Lambda}{2\left(D-1\right)\left(D-2\right)}\right]\; ,
\end{eqnarray}
where $l=0,\ 1,\ 2,\ ...$ for the scalar spherical harmonics. However, the $l=0$ mode is not able to satisfy the gauge condition suggested in the original paper \cite{chm2001}, where the first mode in this case shall start with $l=1$. The dominant term when $r\rightarrow\infty$ is
\begin{eqnarray}\label{vemsa}
V_{s=1,S}\Big|_{r\rightarrow\infty}\sim f\left(r\right)\frac{\left(D-4\right)\left(D-6\right)\Lambda}{2\left(D-1\right)\left(D-2\right)}\sim \frac{\left(D-4\right)\left(D-6\right)\Lambda^{2}r^{2}}{\left(D-1\right)^{2}\left(D-2\right)^{2}} \; ,
\end{eqnarray}
and the other terms in the parentheses of Eq.~(\ref{vems}) are of higher negative orders with ${\cal O}(r^{-a})$ and $a\geq2$. The value of $V_{s=1,S}$ tends to infinity when $r\rightarrow\infty$, excepting for $D=4$ and $D=6$, in which it converges to a finite non-zero value as
\begin{equation}
V_{s=1,S}\Big|_{r\rightarrow\infty,\ D=4,6}\sim \frac{2\Lambda}{\left(D-1\right)\left(D-2\right)}\left[l\left(l+D-3\right)+\frac{\left(D-2\right)\left(D-4\right)}{4}\right]\; .
\end{equation}
It is worth noting that the $D=5$ cases are also special cases where the effective potentials asymptotically diverge to negative infinity, where one can check by Eq.~(\ref{vemsa}). $D=4$,$5$,$6$ are the cases which include {\it non-box-like} potentials for $V_{s=1,S}$.

\subsubsection{The EM-vector perturbation}

\par The effective potential is given by
\begin{eqnarray}\label{vemv}
V_{s=1,V}&=&f\left(r\right)\left[\frac{\left(l+1\right)\left(l+D-4\right)}{r^{2}}+\frac{\left(D-4\right)\left(D-6\right)}{4r^{2}}f\left(r\right)+\frac{\left(D-4\right)}{2r}\frac{df\left(r\right)}{dr}\right]\nonumber\\
&=&f\left(r\right)\left[\frac{l\left(l+D-3\right)}{r^{2}}+\frac{\left(D-2\right)\left(D-4\right)}{4r^{2}}+\frac{D\left(D-4\right)M}{2r^{D-1}}+\frac{\left(D-4\right)\Lambda}{2\left(D-1\right)}\right] \; , \nonumber \\
\end{eqnarray}
where $l=1,\ 2,\ 3,\  ...$ for the vector spherical harmonics. The dominant term when $r\rightarrow\infty$ is
\begin{eqnarray}
V_{s=1,V}\Big|_{r\rightarrow\infty}\sim f\left(r\right)\frac{\left(D-4\right)\Lambda}{2\left(D-1\right)}\sim \frac{\left(D-4\right)\Lambda^{2}r^{2}}{\left(D-1\right)^{2}\left(D-2\right)}\; .
\end{eqnarray}
Similar to the EM-scalar perturbations, as $r\rightarrow\infty$ the $V_{s=1,V}$ in general goes to infinity, except in the $D=4$ cases. The four dimensional asymptotic behavior is given by
\begin{equation}\label{vemvl}
V_{s=1,V}\Big|_{r\rightarrow\infty,\ D=4}\sim \frac{\Lambda l}{3}\left(l+1\right)\; ,
\end{equation}
where the asymptotic behaviors coincide for the EM-scalar and the EM-vector perturbations when $D=4$. Furthermore, the effective potentials are identical in the $D=4$ cases for EM-scalar and EM-vector perturbations.

%
\subsection{The Rarita-Schwinger perturbation}

\par For the Rarita-Schwinger (RS) perturbations (spin-3/2 perturbations), we follow our recent work \cite{ccch2019}. Two types of radial equations were obtained, corresponding to the non-transverse traceless (nTT) and transverse traceless (TT) eigenmodes on $S^{n}$, which have direct analogies to the EM-scalar and EM-vector perturbations in the spin-1 case.

\subsubsection{The RS-nonTT perturbation}

\par The RS-nonTT effective potential in AdS black hole spacetimes is given by
\begin{equation}\label{NTTEFV}
V_{s=3/2,\ nTT}=\mp \partial_{r_{\ast}}\mathcal{W}+\mathcal{W}^{2}\; ,
\end{equation}
where the super potential $\mathcal{W}$ and the tortoise coordinate $dr_{\ast}=\mathcal{F}^{-1}dr$ are
\begin{equation}\label{NTTSP}
\begin{aligned}
   \mathcal{F} = f\left[1 + \frac{f}{2\omega}\left(\frac{\partial}{\partial r}\frac{\mathcal{D}}{i\mathcal{B}} \right)\left(\frac{\mathcal{B}^{2}}{\mathcal{B}^{2}-\mathcal{D}^{2}} \right) \right]^{-1}\; ,\\
    \mathcal{W} = \sqrt{\mathcal{D}^{2}-\mathcal{B}^{2}}\left[1 + \frac{f}{2\omega}\left(\frac{\partial}{\partial r}\frac{\mathcal{D}}{i\mathcal{B}} \right)\left(\frac{\mathcal{B}^{2}}{\mathcal{B}^{2}-\mathcal{D}^{2}} \right) \right]^{-1}\; .
\end{aligned}
\end{equation}
The related coefficient $\mathcal{B}$ and $\mathcal{D}$ are
\begin{eqnarray}\label{NTTCOE}
\mathcal{B} &=& \frac{i\bar{\lambda}\sqrt{f}}{r}\left[1 + \frac{1}{-\bar{\lambda}^{2}+ \frac{(D-2)^{2}}{4}f - r^{2}\frac{\Lambda(D-2)}{2(D-1)}}\left( \frac{(D-2)(D-3)M}{r^{D-3}}\right) \right]\; ,\nonumber\\
\mathcal{D}&=&\sqrt{\frac{\Lambda f (D-2)}{2(D-1)}}\left[\frac{D-4}{D-2} + \frac{1}{-\bar{\lambda}^{2}+ \frac{(D-2)^{2}}{4}f
- r^{2}\frac{\Lambda(D-2)}{2(D-1)}}\left( \frac{(D-2)(D-3)M}{r^{D-3}}\right) \right]\; , \nonumber \\
\end{eqnarray}
where $\bar{\lambda}=j+(D-3)/2$, and $j=3/2,\,5/2,\,7/2,\,...$ is the spinor eigenvalue on $S^{n}$. Note that nTT spinor-vector eigenmodes are the linear combination of the spinor eigenmodes, and it is more convenient to use spinor eigenmodes in the detailed calculation in this case \cite{ccchn2015, ccch2016}. Checking the leading order of $r$ for the parameters in Eqs.~(\ref{NTTEFV}), (\ref{NTTSP}), and (\ref{NTTCOE}) when $r\rightarrow\infty$ we have
\begin{eqnarray}
\mathcal{B}\Big|_{r\rightarrow\infty}\propto r^{0}\; &;& \; \; \mathcal{D}\Big|_{r\rightarrow\infty}\sim\frac{\left(D-4\right)\Lambda r}{\left(D-1\right)\left(D-2\right)};\nonumber\\
\mathcal{W}\Big|_{r\rightarrow\infty}\propto r\; ; \; \; \mathcal{F}\Big|_{r\rightarrow\infty}&\propto&r^{2}\; ; \; \; V_{s=3/2,\ nTT}\Big|_{r\rightarrow\infty}\propto r^{2} \; .
\end{eqnarray}
The effective potential tends to go to infinity when $r\rightarrow\infty$, except for the $D=4$ case. The parameter $\mathcal{D}\Big|_{r\rightarrow\infty,\ D=4}\propto r^{0}$ and the second term in the square parentheses of Eq.~(\ref{NTTSP}) coincidentally vanish, where more explicitly
\begin{equation}
\frac{\mathcal{D}}{i\mathcal{B}}=\frac{2M}{\bar{\lambda}^{3}}\sqrt{\frac{\Lambda}{3}}\; ; \; \; \partial_{r}\frac{\mathcal{D}}{i\mathcal{B}}= 0 \; .
\end{equation}
To summarize these characteristics, the asymptotic behavior of the $V_{s=3/2,\ nTT,\ D=4}$ tends to go to a finite non-zero positive value, and the {\it barrier-like} potential appears in the small black hole area case as show in Ref. \cite{ccch2019}.

\subsubsection{The RS-TT perturbation}

\par The RS-TT effective potential in AdS black hole spacetimes is given by
\begin{equation}\label{TTEFV}
V_{s=3/2,\ TT}=\mp\partial_{{r}_{*}}\mathbb{W}+\mathbb{W}^{2}\; ,
\end{equation}
where the super potential $W$ and the tortoise coordinate $dr_{\ast}=\mathbb{F}^{-1}dr$ are
\begin{eqnarray}\label{TTSP}
\mathbb{F}&=&f\left[1+\frac{f}{2\omega}\sqrt{\frac{\Lambda\left(D-2\right)}{2\left(D-1\right)}}\left(\frac{2\bar{\zeta}\left(D-1\right)}{2\bar{\zeta}^{2}\left(D-1\right)+r^{2} \Lambda\left(D-2\right)}\right)\right]^{-1}\; ,\nonumber \\
\mathbb{W}&=&\left(\frac{\bar{\zeta}^{2}f}{r^{2}}+\frac{f\Lambda\left(D-2\right)}{2\left(D-1\right)}\right)^{\frac{1}{2}} \left[1+\frac{f}{2\omega}\sqrt{\frac{\Lambda\left(D-2\right)}{2\left(D-1\right)}}\left(\frac{2\bar{\zeta}\left(D-1\right)}{2\bar{\zeta}^{2}\left(D-1\right)+r^{2} \Lambda\left(D-2\right)}\right)\right]^{-1}\; ,\nonumber\\
\end{eqnarray}
and the spinor-vector TT eigenvalue is given by $\bar{\zeta}=j+(D-3)/2$, $j=3/2,\,5/2,\,7/2,\,...$. The asymptotic behavior when $r\rightarrow\infty$ is
\begin{equation}
\mathbb{F}\Big|_{r\rightarrow\infty}\propto r^{2}\; ; \; \; \mathbb{W}\Big|_{r\rightarrow\infty}\propto r\; ; \; \; V_{s=3/2,\ TT}\Big|_{r\rightarrow\infty}\propto r^{2} \; .
\end{equation}
The asymptotic behavior tends to go to infinity in this case, as is the general expectation of AdS black holes.

%
\subsection{The Gravitational perturbation}

\par For the Gravitational perturbations we follow the work by Kodama and Ishibashi \cite{ki2003}.

\subsubsection{The Grav.-scalar perturbation}

\par The effective potential is given by
\begin{eqnarray}
V_{s=2,\ S}=\frac{f\left(r\right)Q\left(r\right)}{16r^{2}H^{2}\left(r\right)}\ \ &,& \ \ H\left(r\right)=m+\frac{1}{2}\left(D-1\right)\left(D-2\right)x, \nonumber\\ m=k^{2}_{S}-\left(D-2\right)\ \ &,& \ \ x=\frac{2M}{r^{D-3}},\nonumber\\
k^{2}_{S}=l\left(l+D-3\right)\ \ &,& \ \ l=2,\ 3,\ 4,\ ...,
\end{eqnarray}
where
\begin{eqnarray}\label{Q}
Q\left(r\right)&=&\left[n^{3}\left(n+2\right)\left(n+1\right)^{2}x^{2}-12n^{2}\left(n+1\right)\left(n-2\right)mx+4\left(n-2\right)\left(n-4\right)m^{2}\right]y\nonumber\\
&&+n^{4}\left(n+1\right)^{2}x^{3}+n\left(n+1\right)\left[4\left(2n^{2}-3n+4\right)m+n\left(n-2\right)\left(n-4\right)\left(n+1\right)\right]x^{2}\nonumber\\
&&-12n\left[\left(n-4\right)m+n\left(n+1\right)\left(n-2\right)\right]mx+16m^{3}+4n\left(n+2\right)m^{2}\; .
\end{eqnarray}
and
\begin{eqnarray}
y=\frac{2\Lambda r^{2}}{n\left(n+1\right)}\; , \; \; n=D-2 \; .
\end{eqnarray}
The dominant terms for $V_{s=2,\ S}$ and $Q\left(r\right)$ when $r\rightarrow\infty$ are
\begin{eqnarray}
Q\left(r\right)\Big|_{r\rightarrow\infty}\sim 4\left(n-2\right)\left(n-4\right)m^{2}y\propto r^{2}\; , \; \; V_{s=2,\ S}\Big |_{r\rightarrow\infty}\propto r^{2}\; .
\end{eqnarray}
It is clear that the effective potential goes to infinity when $r\rightarrow\infty$, except for the $D=4$, and $D=6$ cases. For $D=4$, the second term of the first line in Eq.~(\ref{Q}) is proportional to $r$ and will dominate the asymptotic behavior if the ${\cal O} (r^{2})$ term vanishes. However, this term also vanishes because of the coefficient $\left(n-2\right)$.

\par For $D=6$ case, the highest order of $V_{s=2,\ S,\ D=6}$ shall be ${\cal O}(r^{0})$ when $r\rightarrow\infty$, and is possible to observe. That is, the asymptotic behavior of the effective potential $V_{s=2,\ S}$ converges to a finite non-zero value.

\par Lastly, similar to the EM-scalar perturbations, the asymptotic behavior for the effective potential diverges to negative infinity when $D=5$.

\par That is, $D=4$, $5$, $6$ are the cases that include {\it non-box-like} effective potentials.

\subsubsection{The Grav.-vector perturbation}

\par The effective potential is given by
\begin{equation}
V_{s=2,\ V}=\frac{f\left(r\right)}{r^{2}}\left[k^{2}_{v}+1+\frac{\left(D-2\right)\left(D-4\right)}{4}-\frac{3\left(D-2\right)^{2}M}{2r^{D-3}}+\frac{\left(D-4\right)\Lambda r^{2}}{2\left(D-1\right)}\right]\; ,
\end{equation}
where $k^{2}_{V}=l\left(l+D-3\right)-1$, and $l=1,\ 2,\ 3,\ ...$. The {\it non-box-like} potential can be obtained when $D=4$.

\subsubsection{The Grav.-tensor perturbation}

\par The effective potential is given by
\begin{equation}
V_{s=2,\ T}=\frac{f\left(r\right)}{r^{2}}\left[k^{2}_{T}+2+\frac{\left(D-2\right)\left(D-4\right)}{4}+\frac{\left(D-2\right)^{2}M}{2r^{D-3}}+\frac{D\Lambda r^{2}}{2\left(D-1\right)}\right]\; ,
\end{equation}
where $k^{2}_{T}=l\left(l+D-3\right)-2$, and $l=1,\ 2,\ 3,\ ...$. No {\it non-box-like} potential in this case.

%
\subsection{Summary of the effective potentials}

\par In the above discussions for the {\it non-box-like} effective potentials, we can sort the behavior into four cases, ({\it 3 non-box-like} and {\it 1 box-like}):
\begin{itemize}
\item[1.] For the Dirac perturbation, the behaviors of the effective potentials are {\it barrier-like} to {\it step-function} like, and the turning point is located in the large black hole area as shown in Fig.~\ref{PlotD}.

\begin{figure}
\begin{subfigure}{0.3\textwidth}
\includegraphics[width=\textwidth]{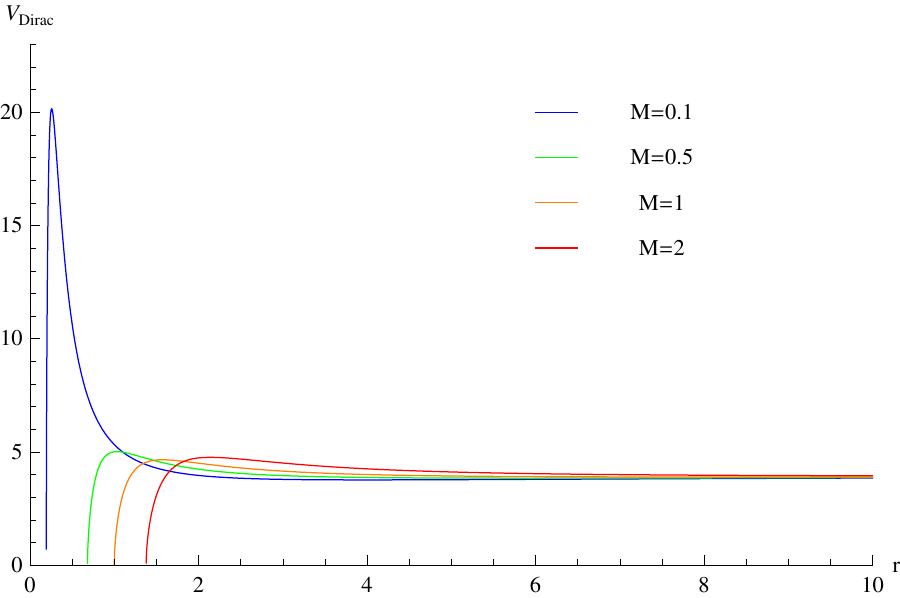}
\caption{$D=4$.}\label{VD4}
\end{subfigure}
\begin{subfigure}{0.3\textwidth}
\includegraphics[width=\textwidth]{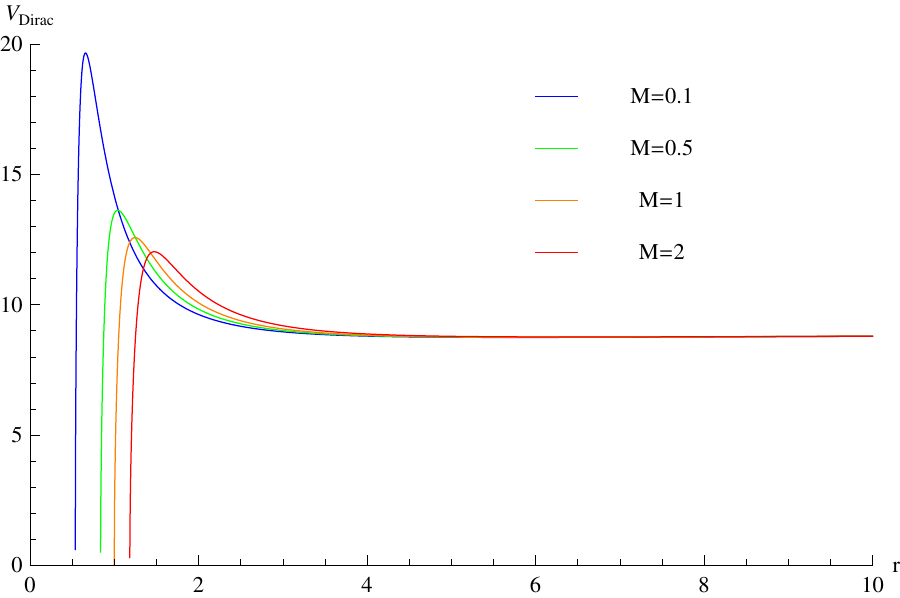}
\caption{$D=6$.}\label{VD6}
\end{subfigure}
\begin{subfigure}{0.3\textwidth}
\includegraphics[width=\textwidth]{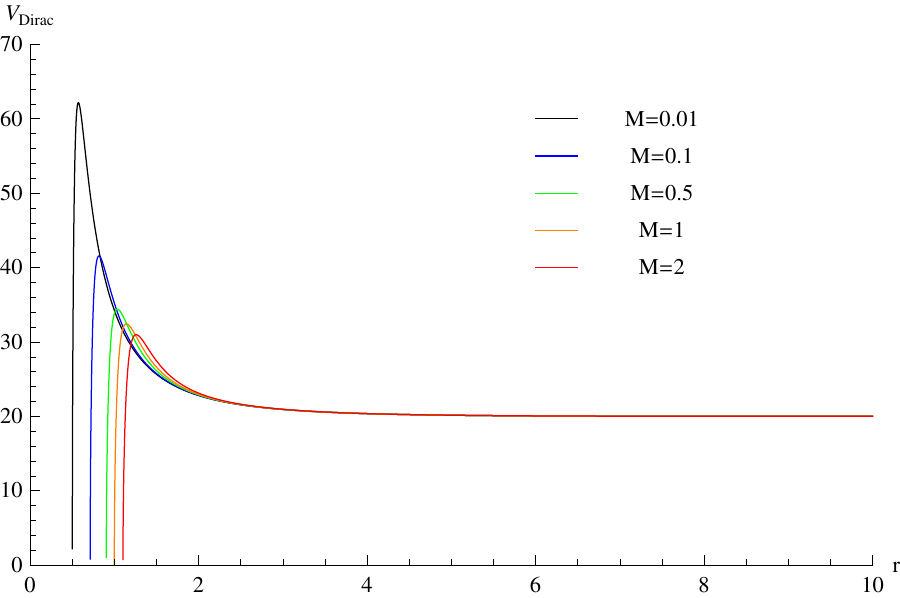}
\caption{$D=9$.}\label{VD9}
\end{subfigure}
\caption{The effective potentials for Dirac perturbations in asymptotic AdS spherically symmetric black holes with $l=1$.}\label{PlotD}
\end{figure}

\item[2.] For the $D=4$ and $6$ EM-scalar and Gravitational-scalar perturbations, $D=4$ EM-vector, RS-nonTT and Gravitational-vector perturbations, the effective potentials are {\it barrier-like} in the small black regions. As the black hole masses increase, the effective potentials approach a {\it step-function}-like shape, and the turning point is located between the small black and the intermediate black hole regions. The behavior is shown in Fig.~\ref{Plotothers}.

\begin{figure}
\begin{subfigure}{0.3\textwidth}
\includegraphics[width=\textwidth]{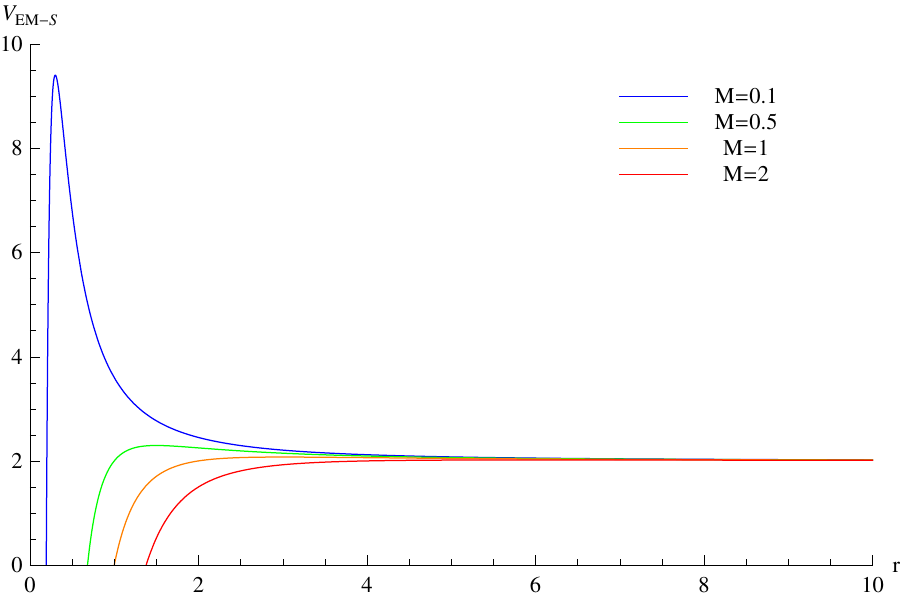}
\caption{The EM-scalar perturbation with $D=4$, $l=1$.}\label{VEMS4}
\end{subfigure}
\begin{subfigure}{0.3\textwidth}
\includegraphics[width=\textwidth]{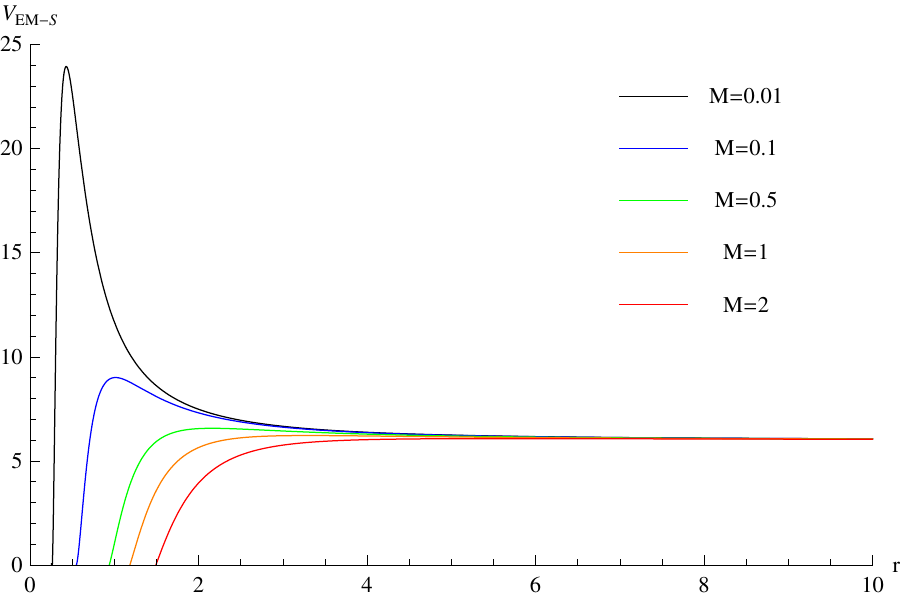}
\caption{The  EM-scalar perturbation with $D=6$, $l=1$.}\label{VEMS6}
\end{subfigure}
\begin{subfigure}{0.3\textwidth}
\includegraphics[width=\textwidth]{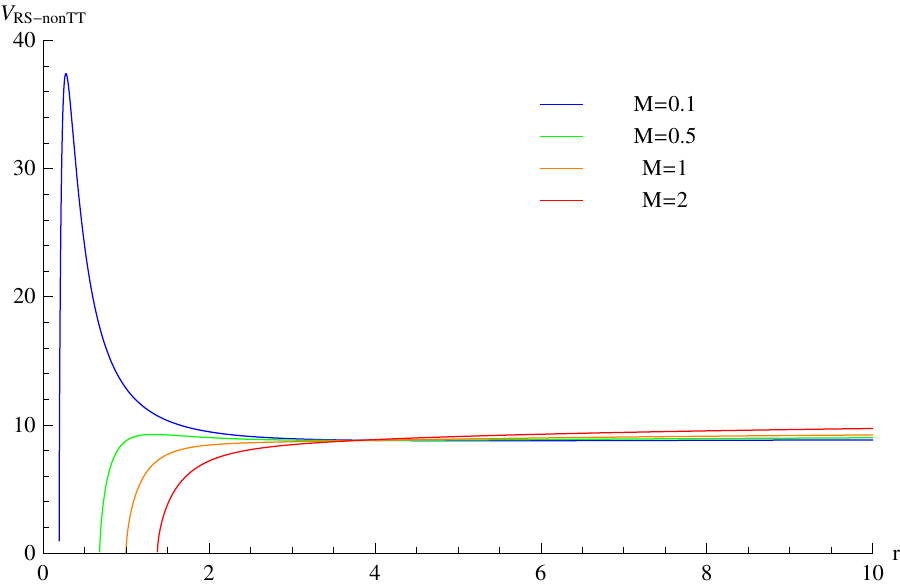}
\caption{The Rarita-Schwinger perturbation with $D=4$, $l=1$.}\label{VRS4}
\end{subfigure}\\
\begin{subfigure}{0.3\textwidth}
\includegraphics[width=\textwidth]{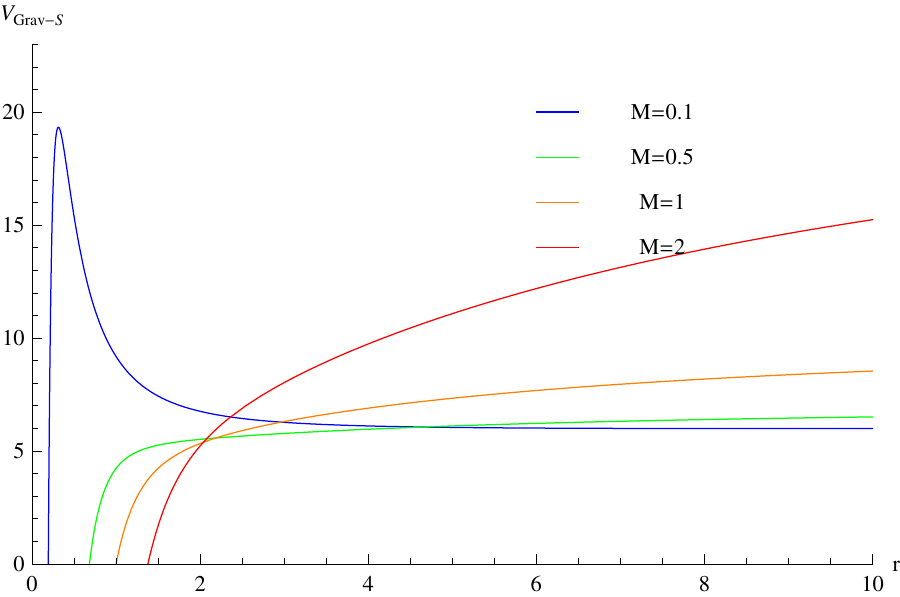}
\caption{The Gravitational-scalar perturbation with $D=4$, $l=2$.}\label{VGS4}
\end{subfigure}
\begin{subfigure}{0.3\textwidth}
\includegraphics[width=\textwidth]{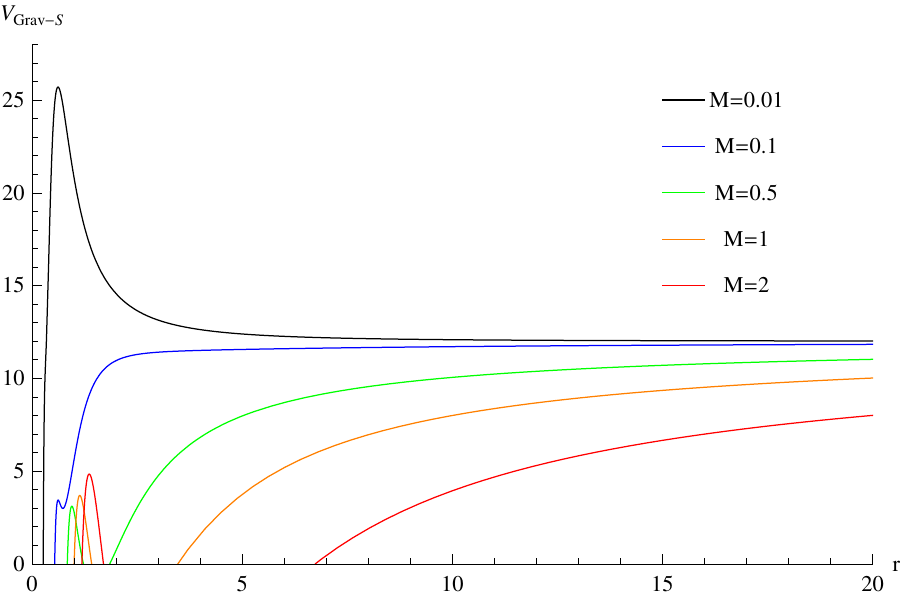}
\caption{The Gravitational-scalar perturbation with $D=6$, $l=2$.}\label{VGS6}
\end{subfigure}
\caption{The cases of the {\it barrier-like} effective potentials for small black holes in asymptotic AdS spherically symmetric spacetimes when $l=2$.}\label{Plotothers}
\end{figure}

\item[3.] For $D=5$, the asymptotic behavior of the effective potentials for EM-scalar and Gravitational-scalar perturbations diverge to negative infinity as shown in Fig.~\ref{PlotEMG5}.
\begin{figure}
\begin{subfigure}{0.3\textwidth}
\includegraphics[width=\textwidth]{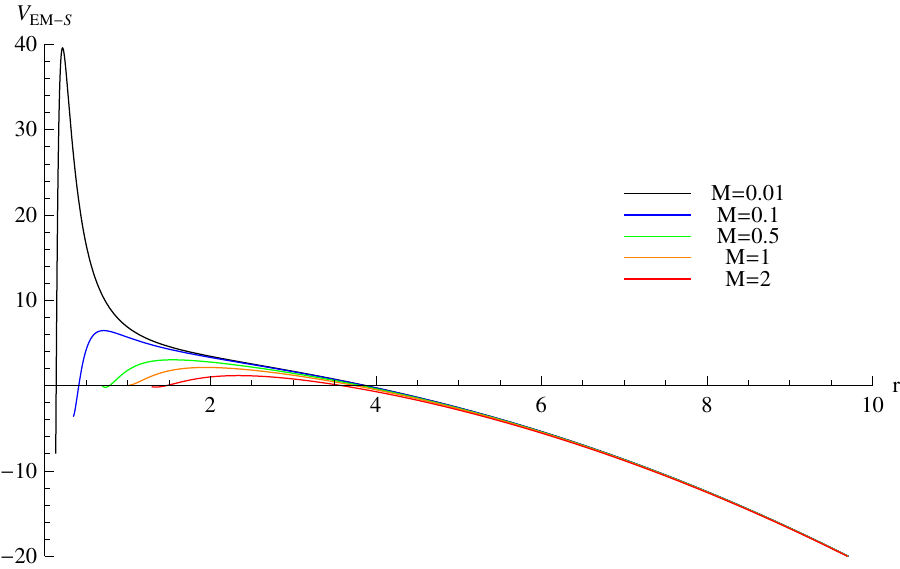}
\caption{The EM-scalar perturbation with $l=1$.}\label{VEMS5}
\end{subfigure}
\begin{subfigure}{0.3\textwidth}
\includegraphics[width=\textwidth]{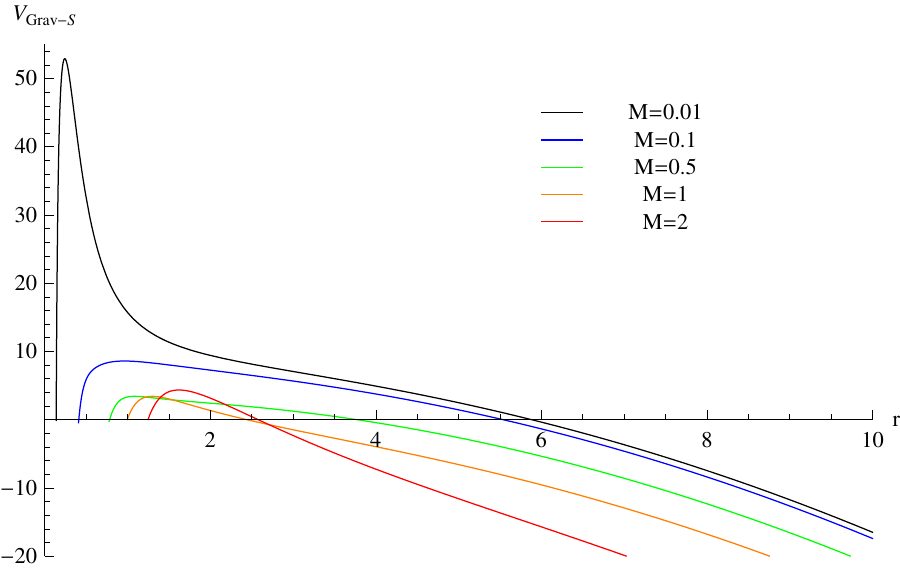}
\caption{The Gravitational-scalar perturbation with $l=2$.}\label{VGS5}
\end{subfigure}
\caption{The effective potentials for EM-scalar and Gravitational-scalar perturbations in 5-dimensional asymptotic AdS spherically symmetric black holes.}\label{PlotEMG5}
\end{figure}

\item[4.] For the scalar perturbations, and the other cases (with the exception to the cases pointed out in this subsection), the effective potentials were identically {\it box-like}.
\end{itemize}

\par To conclude this section, and to compare with the work of Ishibashi and Wald \cite{IW2004}, we found that our {\it non-box-like} potentials strongly overlap with their classifications. Every bosonic case in (2) above, including in specific dimensions, has the asymptotic behavior in the effective potential which is exactly of type (ii) in their classification with the parameter $\nu^{2}=1/4$ (in their notation). For our fermonic cases in (1) and spin-3/2 non-TT case in (2), one can see that the asymptotic behavior for the effective potentials is highly similar to the bosonic cases of Ref. \cite{IW2004}, as such we treat them the same as the other bosonic cases with analogy to the type (ii) case there. For our cases in (3), which overlap with the type (iii) in their work, even though there exists a barrier near the black hole horizon, the effective potentials asymptotically diverge to negative infinity on the AdS boundary. Note that in all the above cases one is allowed to impose a more general Robin boundary condition than the Dirichlet boundary condition in their study. Furthermore, for our cases in (4), which shall be the type (i) in their work with a unique self-adjoint extension, we are not allowed to impose the boundary conditions we consider here as the effective potential is {\it box-like}. Additionally, we do not have the case analogous with their type (iv) as we only consider the massless field in this current work.

\par However, it is important to note that even though the asymptotic behaviors of our potentials fit exactly with the classification of Ishibashi and Wald's work, there exist some intrinsic differences with our work here, including the physical phenomena and the background spacetimes. The self-adjoint extended dynamical solutions in purely AdS spacetimes were their focus, and a new set of QNMs in $SAdS_{D}$ is ours. As a consequence, the constraint equations for the boundary conditions in their work may not fit exactly with those in our studies, and this was also commented on in their conclusion section. However, a common argument is that in such cases a more general Robin boundary condition, where the outgoing wave boundary condition we use falls under this category, is allowed to be used beyond the {\it box-like} boundary conditions.

%
%
\section{Quasi-normal modes}

\par In the previous section we have demonstrated cases where there exist {\it barrier-like} effective potentials in asymptotic AdS black hole spacetimes. From this, the behavior of the related particle interactions with the black hole spacetimes seems to dominate the shape of these effective potentials. Note that the previously studied shape of the effective potentials for AdS black hole spacetimes had strict boundary conditions (b.c.) imposed upon them, which meant a {\it box-like} shape was used for the calculation of QNMs. These conditions were the Dirichlet and the vanishing energy flux (VEF) b.c.'s, where the QNMs obtained from these b.c. had linear relations to the black hole Hawking temperature with variations of the black hole mass. However, due to the {\it barrier-like} behavior of the effective potentials listed in the previous section, it is more natural to impose the usual b.c. when studying {\it barrier-like} effective potentials, that is, that we have purely ingoing waves at the event horizon and purely outgoing waves at  radial infinity:
\begin{eqnarray}
\Psi_{s}&\rightarrow& e^{-i\omega r_{*}}\ \ {\rm as}\ \ r_{*}\rightarrow -\infty \; ,\nonumber\\
\Psi_{s}&\rightarrow& e^{i\sqrt{\omega^{2}-V_{s}|_{r\rightarrow\infty}}\ r_{*}}\ \ {\rm as}\ \ r_{*}\rightarrow 0\ \ ({\rm or}\ \ r\rightarrow\infty) \; .
\end{eqnarray}
The use of these ``usual" b.c. on {\it barrier-like} effective potentials in AdS black hole space times is the main point we would like to highlight in this paper.

\par In the following studies we have chosen the location of the black hole event horizon at $r_{+}=0.2$, to represent small black holes, and the AdS radius fixed to be $1$. This parameter choice is taken mostly for convenience, so as to allow us to compare with the existing results, and show the new set of QNMs. In these cases we shall follow the WKB method \cite{IW1987, Kono2003} and revised WKB methods with Pad\`e approximants \cite{KZZ2019} to obtain the low-lying QNMs. Note that the WKB methods are well suited to studying the {\it barrier-like} effective potentials, and the revised WKB method improves the accuracy over the traditional WKB methods in many ways. Since determining the efficiency of, and finding improved methods for, calculating the QNMs were not the main point of this work, we shall not present the explicit formulae here, but refer the reader to the original papers.

\par In Tab.~\ref{TabD4} and Tab.~\ref{TabD6} we show the low-lying modes for Dirac perturbations when $D=4$ and $D=6$ respectively. In the $D=4$ case we have reference modes for Dirichlet b.c. in \cite{gj2005} and VEF b.c. in Ref. \cite{whj2017}. For the Dirichlet b.c., the authors declared that the Dirichlet b.c. was not sufficient in the small black hole cases and imposed the VEF b.c. a few years later. We list one of the QNM solutions with VEF b.c. in Tab.~\ref{TabD4} as the reference mode, and show a different set of low-lying modes in the same table. For the $D=6$ case we can not find any reference modes with Dirichlet or VEF b.c.'s, and just list our low-lying results in Tab.~\ref{TabD6}.

\begin{table}
\centering
\caption{Low-lying QNMs for Dirac perturbation with $r_{+}=0.2$, $D=4$.}
\begin{tabular}{| c | c | c | c || c | c | }
\hline
 & \multicolumn{3}{|c||}{QNMs }& Reference modes \\
\hline
$\left(l,n\right)$ & 3rd WKB & 6th WKB  & Revised 12th WKB & VEF B.C. I \cite{whj2017}\\
\hline
$\left(0,0\right)$ &1.90051 - 1.11335 i  & 1.96924 - 1.04970 i  & 2.04278 - 1.12047 i &  1.4124 - 1.6293$\times10^{-2}$ i\\
$\left(1,0\right)$ &4.12161 - 1.05958 i  & 4.13981 - 1.05708 i  & 4.13931 - 1.05776 i &  2.4481 - 4.2096$\times10^{-4}$ i\\
$\left(1,1\right)$ &3.79812 - 3.32388 i  & 3.82183 - 3.29897 i  & 3.82709 - 3.30339 i & \\
$\left(2,0\right)$ &6.25938 - 1.05473 i  & 6.26459 - 1.05452 i  & 6.26459 - 1.05449 i & \\
$\left(2,1\right)$ &6.02260 - 3.23251 i  & 6.03369 - 3.22705 i  & 6.03413 - 3.22731 i & \\
$\left(2,2\right)$ &5.65638 - 5.53728 i  & 5.63892 - 5.55768 i  & 5.65012 - 5.5574 i  & \\
\hline
\end{tabular}
\label{TabD4}
\end{table}

\begin{table}
\centering
\caption{Low-lying QNMs for Dirac perturbation with $r_{+}=0.2$, $D=6$.}
\begin{tabular}{| c | c | c | c |  }
\hline
 & \multicolumn{3}{|c|}{QNMs }  \\
\hline
$\left(l,n\right)$ & 3rd WKB & 6th WKB  & Revised 12th WKB \\
\hline
$\left(0,0\right)$ &5.16694 - 2.63149 i  & 5.87845 - 2.33994 i  & 5.6231 - 2.60811 i   \\
$\left(1,0\right)$ &8.55379 - 2.56868 i  & 8.71442 - 2.63564 i  & 8.72762 - 2.59476 i  \\
$\left(1,1\right)$ &6.78673 - 8.48736 i  & 7.27333 - 8.23936 i  & 7.2035 - 8.25834 i   \\
$\left(2,0\right)$ &11.70801 - 2.58225 i & 11.76480 - 2.60582 i & 11.775 - 2.59154 i   \\
$\left(2,1\right)$ &10.38623 - 8.09486 i & 10.59694 - 8.10347 i & 10.6106 - 8.03006 i  \\
$\left(2,2\right)$ &8.293976 - 14.2326 i & 8.39811 - 14.48178 i & 8.47798 - 14.4205 i  \\
\hline
\end{tabular}
\label{TabD6}
\end{table}

\par For the EM-perturbations, we list our low-lying results in Tab.~\ref{TabEMSV4} with $D=4$ and Tab.~\ref{TabEMS6} with $D=6$. When $D=4$ the EM-scalar and EM-vector perturbations are mentioned as even and odd perturbations in some of the literature, and converge to the same effective potential. The reference modes for Dirichlet b.c. \cite{ckl2003} and VEF b.c. \cite{whs2015} are also list in Tab.~\ref{TabEMSV4}. Obviously, our results are a new set of QNMs, and the absolute value of the imaginary part is larger than the previous results. For the $D=6$ case, we also are not able to find any reference modes but just list our low-lying results in Tab.~\ref{TabEMS6}.

\begin{table}
\centering
\caption{Low-lying QNMs for EM-scalar and EM-vector perturbation with $r_{+}=0.2$, $D=4$.}
\begin{tabular}{| c | c | c | c || c | c | }
\hline
 & \multicolumn{3}{|c||}{QNMs }&\multicolumn{2}{|c|}{Reference modes}  \\
\hline
$\left(l,n\right)$ & 3rd WKB & 6th WKB  & Revised 12th WKB & Dirichlet B.C.\cite{ckl2003}& VEF B.C.I\cite{whs2015}\\
\hline
$\left(1,0\right)$ &2.66148 - 1.01454 i  & 2.69097 - 1.00852 i  & 2.69191 - 1.00656 i & 2.63842 - 0.05795 i& 2.6384 - 5.7947$\times10^{-2}$i\\
$\left(1,1\right)$ &2.22718 - 3.30612 i  & 2.23480 - 3.28261 i  & 2.26455 - 3.26632 i & 3.99070 - 0.47770 i&\\
$\left(2,0\right)$ &4.98281 - 1.03784 i  & 4.98870 - 1.03714 i  & 4.98874 - 1.03706 i & &\\
$\left(2,1\right)$ &4.69587 - 3.21313 i  & 4.70441 - 3.20618 i  & 4.70621 - 3.20639 i & &\\
$\left(2,2\right)$ &4.28423 - 5.53861 i  & 4.23772 - 5.60206 i  & 4.32375 - 5.67935 i & &\\
\hline
\end{tabular}
\label{TabEMSV4}
\end{table}

\begin{table}
\centering
\caption{Low-lying QNMs for EM-scalar perturbation with $r_{+}=0.2$, $D=6$.}
\begin{tabular}{| c | c | c | c | }
\hline
 & \multicolumn{3}{|c|}{QNMs }  \\
\hline
$\left(l,n\right)$ & 3rd WKB & 6th WKB  & Revised 12th WKB \\
\hline
$\left(1,0\right)$ &5.49590 - 2.47472 i  & 5.43241 -  2.61125 i  & 5.52868 - 2.49262 i  \\
$\left(1,1\right)$ &3.49387 - 8.09739 i  & 3.09553 -  8.89869 i  & 3.09761 - 4.20542 i  \\
$\left(2,0\right)$ &8.97793 - 2.44011 i  & 8.94638 -  2.42409 i  & 8.97684 - 2.40987 i  \\
$\left(2,1\right)$ &7.49410 - 7.69721 i  & 7.31259 -  7.65382 i  & 7.25787 - 7.39126 i  \\
$\left(2,2\right)$ &5.10392 - 13.5869 i  & 4.20916 - 14.27902 i  & 3.81973 - 14.296 i   \\
\hline
\end{tabular}
\label{TabEMS6}
\end{table}

\par For the spin-3/2 perturbations we show the low-lying results in Tab.~\ref{TabRS4}. In Tab.~\ref{TabGS4} and Tab.~\ref{TabGS6} we present the low-lying QNMs for gravitational-scalar perturbations in $4$- and $6$- dimensional spacetimes, and in Tab.~\ref{TabGV4} we present the low-lying results for gravitational-vector perturbations in the $D=4$ case. For the gravitational-scalar and gravitational-vector perturbations with $D=4$, we have the reference modes from Ref. \cite{ckl2003} evaluated using the Dirichlet b.c.. It is obvious that the QNMs we obtain include a larger absolute value of the imaginary part when comparing with the reference one, which is similar to the EM and Dirac perturbation cases. At the same time, when we fixed $l=2$ and consider the modes with $n=0,1,2$, our results follow the standard behavior of {\it barrier-like} QNMs, where as the mode number is increased, the real part decreases and the absolute value of the imaginary part increases. This is very different from the reference modes, where the real part and the absolute value of the imaginary part both increase (except the first mode of the Grav.-vector perturbation). For the $D=6$ cases, our calculations with the WKB and revised WKB include some results with larger error estimations, even with the revised WKB the number does not conform to the standard expectation for the QNMs behavior in {\it barrier-like} potentials well, though the {\it barrier-like} behavior can be confirmed from Fig.~\ref{VGS6}. Since the applicability of the improved WKB method to this case is beyond the current scope of this work we have not included these modes in the table, though it may be worth noting that this is most likely due to the mathematical form of the Grav.-scalar perturbations, which are much more complicated in this case and required further study with new methods as in \cite{YH2020}.

\par Lastly, the gravitational-scalar perturbations and gravitational-vector perturbations were expected to be iso-spectral for the perturbations in asymptotic flat cases in 4-dimensions. For the asymptotically AdS case we would like to highlight that the effective potentials of these perturbations are still supersymmetric partners, as shown in \cite{MN2002,IB2009}. However, the outgoing wave boundary condition we have adopted does not respect this symmetry, and the QNMs are no longer iso-spectral. This can be seen from our results in Tab.~\ref{TabGS4} and Tab.~\ref{TabGV4}. This breaking of symmetry by the boundary condition was discussed in some detail in Refs. \cite{MN2002,IB2009}, in which both the Dirichlet b.c. and the VEF b.c. were used in these studies. Note that in order to respect this symmetry it is necessary to implement a mixed type Robin boundary condition, and not the Dirichlet or VEF b.c.'s. Furthermore, it should be noted that the outgoing wave boundary condition does respect this symmetry in the asymptotically flat case.

\begin{table}
\centering
\caption{Low-lying QNMs for RS-nonTT perturbation with $r_{+}=0.2$, $D=4$.}
\begin{tabular}{| c | c | c | c | }
\hline
 & \multicolumn{3}{|c|}{QNMs }  \\
\hline
$\left(l,n\right)$ & 3rd WKB & 6th WKB  & Revised 12th WKB \\
\hline
$\left(1,0\right)$ &5.77663 - 1.01874 i  & 5.78311 - 1.01859 i  & 5.78313 - 1.01852 i  \\
$\left(1,1\right)$ &5.50969 - 3.13705 i  & 5.52465 - 3.12911 i  & 5.52565 - 3.12932 i  \\
$\left(2,0\right)$ &8.02053 - 1.03376 i  & 8.02282 - 1.03368 i  & 8.02282 - 1.03367 i  \\
$\left(2,1\right)$ &7.82594 - 3.14384 i  & 7.83180 - 3.14158 i  & 7.83193 - 3.14166 i  \\
$\left(2,2\right)$ &7.49593 - 5.35020 i  & 7.48683 - 5.35604 i  & 7.49055 - 5.35699 i  \\
\hline
\end{tabular}
\label{TabRS4}
\end{table}

\begin{table}
\centering
\caption{Low-lying QNMs for gravitational-scalar perturbation with $r_{+}=0.2$, $D=4$.}
\begin{tabular}{| c | c | c | c || c | }
\hline
 & \multicolumn{3}{|c||}{QNMs }&  Reference modes \\
\hline
$\left(l,n\right)$ & 3rd WKB & 6th WKB  & Revised 12th WKB & Dirichlet B.C.\cite{ckl2003}\\
\hline
$\left(2,0\right)$ &4.07298 - 0.95156 i  & 4.08240 - 0.94900 i  & 4.08182 - 0.94964 i & 3.56571 - 0.01432 i \\
$\left(2,1\right)$ &3.67544 - 3.00497 i  & 3.68710 - 2.98361 i  & 3.69847 - 2.98371 i & 4.83170 - 0.26470 i\\
$\left(2,2\right)$ &3.12784 - 5.26440 i  & 2.93923 - 5.41662 i  & 3.14441 - 5.27870 i & 6.17832 - 0.82063 i\\
$\left(3,0\right)$ &6.54307 - 1.00323 i  & 6.54536 - 1.00289 i  & 6.54536 - 1.00290 i & \\
$\left(3,1\right)$ &6.30285 - 3.07144 i  & 6.30746 - 3.06755 i  & 6.30802 - 3.06794 i & \\
$\left(3,2\right)$ &5.91577 - 5.26597 i  & 5.88580 - 5.28130 i  & 5.89971 - 5.28231 i & \\
$\left(3,3\right)$ &5.46208 - 7.55698 i  & 5.31998 - 7.71147 i  & 5.41247 - 7.68064 i & \\
\hline
\end{tabular}
\label{TabGS4}
\end{table}

\begin{table}
\centering
\caption{Low-lying QNMs for gravitational-scalar perturbation with $r_{+}=0.2$, $D=6$.}
\begin{tabular}{| c | c | c | c | }
\hline
 & \multicolumn{3}{|c|}{QNMs}  \\
\hline
$\left(l,n\right)$ & 3rd WKB & 6th WKB  & Revised 12th WKB \\
\hline
$\left(2,0\right)$ &6.01307 - 1.90693 i  & 6.86528 - 1.79422 i  & 6.05701 - 1.84091 i  \\
$\left(2,1\right)$ &4.67896 - 5.98387 i  &   & 4.44256 - 5.93977 i  \\
$\left(2,2\right)$ &2.38796 - 10.60592 i &   & 2.02932 - 11.90730 i \\
$\left(3,0\right)$ &10.02829 - 2.03815 i & 9.84558 - 1.97446 i  & 10.03650 - 2.05255 i \\
$\left(3,1\right)$ &8.83686 - 6.27198 i  & 7.77432 - 6.15108 i  & 8.85268 - 6.37837 i  \\
$\left(3,2\right)$ &6.62631 - 10.97617 i &   & 6.06546 - 11.39350 i  \\
$\left(3,3\right)$ &3.73178 - 16.23193 i & &\\
\hline
\end{tabular}
\label{TabGS6}
\end{table}

\begin{table}
\centering
\caption{Low-lying QNMs for gravitational-vector perturbation with $r_{+}=0.2$, $D=4$.}
\begin{tabular}{| c | c | c | c || c | }
\hline
 & \multicolumn{3}{|c||}{QNMs }& Reference modes  \\
\hline
$\left(l,n\right)$ & 3rd WKB & 6th WKB  & Revised 12th WKB & Dirichlet B.C.\cite{ckl2003}\\
\hline
$\left(2,0\right)$ &4.07553 -  0.95336 i  & 4.08045 - 0.94898 i  & 4.08143 - 0.94977 i & 2.404 - 3.033 i\\
$\left(2,1\right)$ &3.69053 -  3.00948 i  & 3.67890 - 2.97921 i  & 3.68819 - 2.98674 i & 4.91594 - 0.30408 i\\
$\left(2,2\right)$ &3.16662 -  5.26691 i  & 2.91197 - 5.40282 i  & 3.12443 - 5.25329 i & 6.30329 - 0.89773 i\\
$\left(3,0\right)$ &6.54309 -  1.00324 i  & 6.54536 - 1.00289 i  & 6.54536 - 1.0029 i & \\
$\left(3,1\right)$ &6.30293 -  3.07148 i  & 6.30746 - 3.06751 i  & 6.30799 - 3.06796 i & \\
$\left(3,2\right)$ &5.91600 -  5.26602 i  & 5.88576 - 5.28114 i  & 5.89992 - 5.28302 i & \\
$\left(3,3\right)$ &5.46255 -  7.55704 i  & 5.31979 - 7.71105 i  & 5.41502 - 7.68205 i & \\
\hline
\end{tabular}
\label{TabGV4}
\end{table}

%
%
\section{Conclusion}

\par The main results of this work has been to succinctly collect a set of non {\it box-like} effective potentials for the massless perturbations of spherically symmetric AdS black hole spacetimes, and obtain the corresponding low-lying QNMs for the {\it barrier-like} cases. The new (original) set of low-lying QNMs follow the standard behaviors of {\it barrier-like} cases, where when fixing the angular parameter, and when the mode number increase, the real part decreases and the absolute value of imaginary part increases. This is obviously a different set of QNMs to those obtained by the Horowitz and Hubeny methods with Dirichlet or VEF b.c.'s. With these results we shall summarize some future and open questions as follows:
\begin{itemize}
\item[1.] In String theory and the AdS/CFT correspondence, the spacetime is estimated to be $S^{m} \times AdS_{D}$ with various types of models with $D=4$, $5$, and $6$. The phenomena of QNMs obtained by the Horowitz and Hubeny methods with Dirichlet b.c. were expected to be related to the poles of retarded Green's functions under the AdS/CFT correspondence. The new set of QNMs obtained in this paper corresponds, under the AdS/CFT correspondence, to an interesting new category of phenomena, as suggested by the different b.c.'s in Refs. \cite{IW2004, CM2008, H2009, AST2009}.
\item[2.] In cosmological observations the gravitational QNMs in asymptotic flat or dS spacetimes were expected to be dominated by the ring-down behavior of the gravitational waves. On the theoretical sides, this type of QNM is dominated by the {\it barrier-like} effective potentials, which are similar to the cases here. The difference is that our background spacetimes were AdS, where the cosmological constant is negative. This is not consistent with the observed results, which require positive, small and non-zero cosmological constants. It is, however, within schemes such as String theory, that AdS spacetimes still survive, especially for AdS$_{4,5,6}$, in which the non {\it box-like} cases illustrated in this paper are particularly prevalent. We expect that these results can be a tested in cosmological AdS spacetimes with extra dimensions, for example, a brane model with an $SAdS_{D}$ bulk.
\item[3.] The non {\it box-like} effective potentials may just exist for the massless perturbations of spherically symmetric AdS black hole spacetimes. At least for the massive Dirac and EM perturbation cases, the effective potentials are {\it box-like} as they diverge in the asymptotic limit when the radial coordinate goes to infinity also in spherically symmetric AdS black hole spacetimes. Similarly, it would be interesting to see how the outgoing wave boundary condition could be generalized to the AdS rotational black hole cases \cite{GM2005}.
\item[4.] The effective potential can not only calculate the QNMs, but also the transmission probabilities, which are very different between {\it barrier-like} potentials and {\it box-like} potentials. The implication for this direction of consideration is for the corresponding grey-body factors, absorption cross-sections, and Hawking radiation. The phenomena in asymptotic AdS spacetimes can also be studied on the microscopic scale. As a remark for microscopic black holes, when we consider the {\it barrier-like} effective potential in the extremely small black hole limit, the behavior will approach a delta-function-like barrier.

\end{itemize}

%
%

\section*{Acknowledgements}

ASC is supported in part by the National Research Foundation of South Africa (NRF).  HTC is supported in part by the Ministry of Science and Technology, Taiwan, under the Grants No. MOST108-2112-M-032-002.

%
%

\end{document}